\documentclass[aps,preprint,nofootinbib]{revtex4}
\usepackage{xcolor}
\usepackage{amsfonts}
\usepackage{amssymb}
\usepackage{amsmath}
\usepackage[font={footnotesize,it}]{caption}
\usepackage{latexsym}
\usepackage{changes}
\usepackage{hyperref}
\usepackage[capitalise]{cleveref}

\usepackage[caption = false]{subfig}
\usepackage{graphicx}
\usepackage{float}
\usepackage[font=small,labelfont=bf,
   justification=justified,
   format=plain]{caption}
\hypersetup{
    colorlinks=true,
    linkcolor=blue,
    filecolor=magenta, citecolor=blue,    urlcolor=blue,
}
\begin{document}
\title{Probing naked singularities in the charged and uncharged $\gamma - metrics$ with quantum wave packets }
\author{Ozay Gurtug \footnote{Corresponding Author}}
\email{ozaygurtug@maltepe.edu.tr}

\affiliation{T. C. Maltepe University, Faculty of Engineering and Natural Sciences,
34857, Istanbul -Turkey}
\author{Mustafa Halilsoy}
\email{mustafa.halilsoy@emu.edu.tr}
\affiliation{Department of Physics, Faculty of Arts and Sciences, Eastern Mediterranean
University, Famagusta, North Cyprus via Mersin 10, Turkey}
\author{Mert Mangut}
\email{mert.mangut@emu.edu.tr}
\affiliation{Department of Physics, Faculty of Arts and Sciences, Eastern Mediterranean
University, Famagusta, North Cyprus via Mersin 10, Turkey}

\begin{abstract}
The non-trivial naked singularities that possess directional behavior in the charged and uncharged Zipoy-Voorhees (ZV) spacetimes, known as $ \gamma - metrics $ are investigated within the context of quantum mechanics. Classically singular spacetime is understood as a geodesic incompleteness with respect to a particle probe, while quantum singularity is understood as a non-unique evolution of test quantum wave packets. In this study, quantum wave packets obeying Klein - Gordon equation are used to probe timelike naked singularities. It is shown by rigorous mathematical calculations that the outermost singularity developed in the charged and uncharged Zipoy-Voorhees spacetime on the equatorial plane is quantum mechanically singular for all values of the deformation parameter $\gamma$. However, directional singularities that develop on the symmetry axis is shown to be healed partially for specific range of the  parameter $\gamma$, if the analysis is restricted purposely to only specific mode (s-wave mode). Allowing arbitrary modes, classical directional singularities remains quantum singular.

\end{abstract}
\maketitle
\section{Introduction}
The Kerr metric is accepted to be the most important solution to  Einstein's field equations, which describes the field of a rotating black hole. It is generated from a spinning spherical object and this unique feature makes Kerr metric a very convenient choice for analyzing idealized astrophysical compact objects.

However, observations reveal that astrophysical compact objects are not perfectly spherical. Our planet is a good example of this fact. In view of this, understanding the observable universe in terms of the mass quadrupole moment of non-spherical astrophysical compact objects requires  understanding the static axially symmetric solutions of Einstein's equations that possess a mass quadrupole term. Among the others, an intriguing  solution of Einstein's equations that describes the geometry of non-spherical compact objects is the static axially-symmetric solution, which is known as the $\gamma-$ metric or Zipoy-Voorhees $(ZV)$ class of metrics \cite{1,2,3}. This metric reduces to static spherically symmetric Schwarzchild solution when  $\gamma=1$. For $0<\gamma<1$, the central compact object is  prolate, whereas for $\gamma>1$, it is  oblate. Hence,  $\gamma$ measures the deviation from  spherical symmetry. It has been known that when $0<\gamma \neq 1$, the resulting solution is horizonless and displays naked singularities at different locations. Some of these singularities show directional properties such that the curvature invariant becomes unbounded at different points when it is approached from different directions. To be more precise, when the singularity is approached along the north pole $ (\theta=0)$ or south pole  $ (\theta=\pi) $, the curvature invariant becomes bounded for $\gamma\geq 2$. This directional behavior of the singularity has been analysed classically in \cite{4}, by employing different coordinate systems.  The main motivation of authors was to understand this behavior, however, their attempt was unsuccessful. In another study \cite{5}, it has been shown that the naked singularities of $ZV$ metrics are point-like for $\gamma<0$, string-like for $0<\gamma<1$ and ring-like when $\gamma>1$. \\
At this stage it is crucial to clarify some points about the types of spacetime singularities in general relativity. A classical singularity is indicated by incomplete geodesics (i.e. there exist spacetime points not accessible by geodesics.) or incomplete paths of bounded acceleration \cite{a,b}. This implies that the future time evolution of a test particle cannot be predicted. Moreover, curvature invariants become unbounded when the singularity is approached. Another type is the quantum singularity. In this type, a spacetime is quantum mechanically nonsingular if the time evolution of a test quantum wave packet is uniquely determined by the initial wave packet. This is achieved in static spacetimes if the spatial portion of the Klein-Gordon operator is essentially self adjoint. Otherwise the spacetime is regarded as quantum mechanically singular. \\

Recently, the charged version of $ZV$ metric is obtained in \cite{6}. The new metric reduces to the spherically symmetric Reissner-Nordstrom solution when $\gamma=1$. However, as in the case of uncharged $ZV$ metric, the solution becomes singular whenever $0<\gamma\neq 1$. The Weyl scalar $\Psi_{2}$ indicates that the new solution also possesses directional singularities whenever $0<\gamma^{2}<3$, except for $\gamma\neq 1$. The interesting case is encountered when the singularity is approached along the symmetry axis, namely, the north pole or the south pole. In this particular case, the Weyl scalar $\Psi_{2}$ remains finite as long as $\gamma\geq 2$.

The main purpose of this paper is two fold. First, the naked singularities that form in the charged and uncharged ZV solutions for all values of $\gamma $ except zero and one will be investigated.  As a second scope, the directional singularities that form along the symmetry axis  will be analyzed separately. Our singularity analysis for both cases will be based on the principles of quantum mechanics.  Our motivation for using a quantum mechanical approach is to investigate the naked singularities with waves rather than particles.

In this article, the method proposed by Wald \cite{7}, which was developed by Horowitz and Marolf (HM)\cite{8}, will be used for analyzing the naked singularities. This method utilizes the unique time evolution of wave packets for all times. A unique time evolution restricts the spatial part of the wave operator to be essentially self-adjoint. And hence, a spacetime is regarded as nonsingular if the evolution of quantum states are uniquely determined for all times with the assigned initial conditions.

Although the method of HM can only be used in static spacetimes with timelike singularities, it has been found powerful and consistent, and thus, has been used in a variety of spacetimes to see if there is any chance to regularize the singular spacetimes with quantum waves rather than with particles \cite{9,10,11,15,24}. Studies so far reveal that the understanding of  spacetime singularities has not been completed yet. Timelike singularities in static spacetimes addressed in this study constitute only one particular type. We still lack a complete understanding of spacelike and null singularities in view of quantum mechanics.

The paper is organized as follows. Section \ref{2} briefly describes  the charged $ZV$ metric found recently in \cite{6}. Definitions of quantum singularities and the foregoing analysis are presented in section \ref{3}. The paper is concluded with results and discussions in section \ref{4}.

\section{Brief Review of the charged $ZV-$ Metric }\label{2}

Recently, the charged version of the ZV metric valid for all values of $\gamma $ is obtained in \cite{6}. In obtaining the solution, the metric describing the collision of gravitational waves coupled with electromagnetic waves \cite{C1,C2} is utilised. In this formalism, the solutions to the field equations are found in prolate coordinates, which are useful in the description of colliding gravitational waves. However, the resulting solution is interpreted in a different coordinate system by transforming the solution into the static, non-spherical form of $ZV$. With reference to the paper cited as \cite{6}, the metric describing static, axially symmetric charged non-spherical compact objects is given by

\begin{equation}
ds^{2}=\frac{\Delta^{\gamma}}{K^{2}}dt^{2}-\frac{K^{2}}{\Delta^{\gamma}}\left[ \Delta^{\gamma^{2}}\Sigma^{1-\gamma^{2}}\left( \frac{dr^{2}}{\Delta}+r^{2}d\theta^{2}  \right)+r^{2}\Delta sin^{2}\theta d\varphi^{2}  \right],
\end{equation}
where

 \begin{equation}
\Delta(r)=1-\frac{2m}{r}+\frac{m^{2}q^{2}}{r^{2}}
\end{equation}

\begin{equation}
\Sigma(r,\theta)=1-\frac{2m}{r}+\frac{m^{2}}{r^{2}}\left( q^{2}+p^{2}sin^{2}\theta \right)
\end{equation}

and

\begin{equation}
K=K(r)=(1+p)\left(1-\frac{m(1-p)}{r}   \right)^{\gamma}-(1-p)\left(1-\frac{m(1+p)}{r}   \right)^{\gamma},
\end{equation}

where  parameters $p$ and $q$ are real constants that satisfy $p^{2}+q^{2}=1$. Here, $q$ is a charge related parameter defined by $Q=mq$, and the physical mass is related to $M=m\gamma$ . It is straightforward to check that  solution (1) reduces to Schwarzchild solution with $\gamma=1$, $q=0$ and $p=1$ (followed by a trivial scaling of coordinates). When $\gamma=1$, \textcolor{blue}{$p=1$} and $q\neq0$, solution (1) reduces to the Reissner-Nordstrom.

We propose two electromagnetic fields as possible candidates for the spacetime of concern.

i) Pure magnetic source with the vector potential

\begin{equation}
A_{\mu}=(0,0,0,C_{0}cos\theta),
\end{equation}
where $C_{0}=const.=2\sqrt{2}m\gamma pq$ is a magnetic charge. This corresponds to the energy-momentum tensor, $T_{\mu}^{\nu}=\frac{I}{4}diag(-1,-1,1,1)$ in which $I=\frac{2C^{2}_{0}}{r^{4}K^{4}}\Delta^{-(\gamma-1)^{2}}\Sigma^{\gamma^{2}-1}$ satisfying the Einstein-Maxwell equations $R_{\mu}^{\nu}=-T_{\mu}^{\nu}$ $(8\pi G=1=c).$

ii) Pure electric source with the static potential
\begin{equation}
A_{\mu}=(f(r),0,0,0),
\end{equation}
where $f(r)=C_{0} \int^{r} \frac{\Delta^{\gamma-1}dr}{r^{2}K^{2}}$ with the energy-momentum tensor $T_{\mu}^{\nu}=\frac{I}{4}diag(1,1,-1,-1)$, in which $I=-\frac{2C^{2}_{1}}{r^{4}K^{4}}\Delta^{(\gamma-1)^{2}}\Sigma^{\gamma^{2}-1}$

The singularity structure of charged $ZV-$ metric can be examined by calculating the null tetrad (Weyl scalar), $\Psi_{2}$. With reference to  paper \cite{6}, it is given by

\begin{equation}
\begin{aligned}
\Psi_{2}(r,\theta)&=\frac{\Delta^{1+\gamma-\gamma^{2}}}{4K^{2}}\Sigma^{\gamma^{2}-1}\left\{\frac{m}{r^{4}\Delta}\left[(r-mq^{2})(1-\gamma^{2}-2\gamma)+m\gamma q^{2}(\gamma-1) \right]+\frac{3(1-\gamma^{2})}{r^{2}}   \right.\\
&\left.+\frac{\gamma(\gamma-1)(\gamma-\gamma^{2}-2)m^{2}(r-mq^{2})^{2}}{r^{6}\Delta^{2}}+\frac{(1-\gamma^{2})(\gamma^{2}-2)m^{2}\left[m(sin^{2}\theta+q^{2}cos^{2}\theta)-r \right]^{2}}{r^{6}\Sigma^{2}}
\right.\\
&\left.-\frac{(1-\gamma^{2})(1+2\gamma-2\gamma^{2})m^{2}(mq^{2}-r)\left[r-m(sin^{2}\theta+q^{2}cos^{2}\theta) \right]}{r^{6}\Delta\Sigma}+\frac{(1-\gamma^{2})(4m-3r)}{r^{3}\Sigma}
\right.\\
&\left. +\frac{K^{\prime\prime}}{K}+\frac{K^{\prime}}{K}\left[-3\frac{K^{\prime}}{K}+\frac{2}{r}\left(-\gamma^{2}+\frac{m\gamma(2-\gamma)(r-mq^{2})}{r^{2}\Delta}+\frac{(1-\gamma^{2})(r-m)}{r\Sigma} \right) \right]
\right.\\
&\left.+\frac{(1-\gamma^{2})p^{2}m^{2}}{r^{4}\Sigma^{2}}\left( cos2\theta-\frac{p^{2}m^{2}}{\Delta r^{2}}sin^{2}\theta  \right) \right\}.
\end{aligned}
\end{equation}

We note that in this expression the terms involving $K$, $K^{\prime}=\frac{dK}{dr}$ and $K^{\prime\prime}=\frac{d^{2}K}{dr^{2}}$ are not given explicitly. This is due to the fact that since $K(r)\neq0$ (for $p>0$), we do not get any extra singularities, and $K(r)$ indicates the existence of a Maxwell source. A careful analysis reveals that possible singularities occur at $r=0,$ and the roots of the functions can be found by setting $\Delta(r)=0$ and $\Sigma(r,\theta)=0.$ We find that for $\gamma>0,$ the outermost singularitiy is due to the root of $\Delta(r)=0,$ which implies

\begin{equation}
r_{\Delta}=m(1+p),
\end{equation}
as long as $\gamma\neq1.$ The singularities due to the roots of $\Sigma(r,\theta)=0$ are valid only for $\gamma^{2}<3$. To see this, we make a power counting of $\Sigma(r,\theta)$ in Eq.(7). Consequently, we obtain

\begin{equation}
r_{\Sigma}=m(1+pcos\theta)
\end{equation}
which shows that for $\theta=0,$ it coincides with the root of $\Delta(r)=0.$ Since $0\leq\theta\leq\pi$, the singularity of $\Sigma(r,\theta)=0$ satisfies

\begin{equation}
m(1-p)\leq r_{\Sigma}\leq m(1+p).
\end{equation}

In the equatorial plane ($\theta=\pi/2$) we have $r_{\Sigma}=m$ which corresponds to a ring singularity. The $\theta-$ dependence in the singularity structure is peculiar to all $\gamma-$ metrics, which are called directional singularities. This implies that approaching to each singularity from different directions will give rise to different powers of divergences.
In summary, the charged ZV-metrics for $\gamma>0$, (with the conditions $\gamma\neq1$) admit naked singularities without horizons. The metric displays multi-singular structure at different radial hypersurfaces. The outermost singularity $r_{\Delta}=m(1+p)$ covers all the others, which will be the target of our probe. \\
Since the aim of this study is to analyse timelike naked singularities, it is very important to verify the character of the naked singularities developed in the ZV spacetimes whether it is nulllike or timelike. Simply we can show that the surface described by $S(r)=r-r_+$ for $r_+=m(1+p)$ is timelike by calculating the normal vector to this surface in the limit of $r\rightarrow r_+$:
\begin{equation}
\begin{aligned}
\left(\nabla S \right)^2&=g^{rr}\left(\frac{dS}{dr}\right)^2=-\frac{\Sigma^{\gamma^{2-1}}}{K^2}\Delta^{1+\gamma-\gamma^2} \lvert_{r\rightarrow r_+}\\
&=-\frac{1}{K^2}\left( \frac{m^2p^2sin^2\theta}{r_+^2}\right)^{\gamma^2-1}\left( \frac{(r-r_+)\left(r-r_-\right)}{r_{+}^{2}}\right)^{1+\gamma-\gamma^2}<0,
\end{aligned}
\end{equation}
where $r_-=m(1-p)$ and $K=\left(2p \right)^\gamma\left(1+p \right)^{1-\gamma}<\infty$, in the limit of $r\rightarrow r_+$. Since $\left(\nabla S \right)^2<0,$ Normal vector to singular surface $S$ is spacelike implies that the singular surface $S$ is timelike. This result is valid also for the uncharged case. \\
Another alternative way for determining the character of the singular surface is to draw the corresponding Penrose diagram of that spacetime. The first step is to transform the coordinates $t$ and $r$ in Eq. (1) into a new coordinates $T$ and $R$ such that
\begin{equation}
\begin{aligned}
&T=tan^{-1}(t+r^{*})+tan^{-1}(t-r^{*})\\
&R=tan^{-1}(t+r^{*})-tan^{-1}(t-r^{*}),
\end{aligned}
\end{equation}
where
\begin{equation}
r^{*}=\int\sqrt{\frac{g_{rr}}{g_{tt}}}dr.
\end{equation}
The timelike or nulllike character of the surface can be verified for a finite value of $t$; if $r^{*}\rightarrow -\infty$ for $r\rightarrow r_{+}$, then the singularity at $r=r_{+}$ is nulllike singularity. On the other hand, the singularity at $r=r_+$ is timelike, if $r^*$ is at a finite value when $r\rightarrow r_+$. Conseqvently, the functional form of $r^*(r)$ is crucial to understand the character of the singularity. If we apply this method to the uncharged $ZV$ metric, we have
\begin{equation}
r^*=\int\left( 1-\frac{2m}{r}\right)^{-\gamma}\left(\frac{r^2-2mr}{r^2-2mr+m^2sin^2\theta}\right)^{\frac{\gamma^2-1}{2}}dr.
\end{equation}
The exact analytic solution to this integral is almost hopeless. The solution to this integral is obtained by expanding the integrand near the singular surface $r_+=2m$ and taking only the first (dominant) term which yields
\begin{equation}
r^*\approx \frac{2^{\gamma^2}(r-r_+)^{\frac{\left(\gamma-1 \right)^{2}}{2}}r_+^{\gamma-\frac{\gamma^2}{2}+\frac{1}{2}}sin^{1-\gamma^2}\theta}{\left(\gamma-1 \right)^{2}}.
\end{equation}
In the limit $r\rightarrow r_+$, $r^*<\infty$. This indicates that the singularity at $r=r_+=2m$ is timelike, because the new coordinate $R=0$. Hence, the singularity appears on the vertical axis ($T$- axis) of the Penrose diagram.

Our aim in this paper is to understand the non-trivial examples of naked singularities developed in the charged and uncharged ZV metrics in view of quantum mechanics. In doing so, quantum wave packet obeying the Klein-Gordon equation will be used to probe the singularity.

\section{Quantum Probe of Naked Singularities}\label{3}

Understanding and resolving  spacetime singularities in the solutions of Einstein's equations are of upmost importance as long as the deterministic nature of the theory of classical relativity is concerned. The resolution of singularities becomes more important in cases when the spacetime singularity is not hidden by horizon(s) and becomes visible to distant observers - naked singularities. Classically, singularities are known as a geodesic incompleteness (with respect to point particle probe). An important indication of the existence of a singularity is the presence of unbounded curvature invariants. As a consequence, all known laws of physics become invalid and classical attempts to understand singularities become incapable for further analysis. In such a regime where the curvature invariants become unbounded, it becomes essential to take quantum effects into account. Thus, analysing the singularities in these regimes requires the use of the tools of quantum theory of gravity. Unfortunately, a consistent quantum theory of gravity has not been developed yet. In view of this fact, any method that incorporates  quantum mechanics in analysing the singularity can be regarded as a right step.

One of the important contributions along this line is the physically sensible prescription proposed by Wald \cite{7}. In this prescription, a Klein-Gordon massless scalar field propagating in an arbitrary static spacetime admitting a timelike singularity can be used to probe the singularity. Thus, the classical particle probe is replaced by a quantum wave packet (scalar wave obeying Klein-Gordon equation). Here, the evolution of a quantum wave packet is translated into the problem of defining self-adjoint extension of the spatial part of the wave operator. This idea is further developed by Horowitz and Marolf \cite{8}, who proposed that a classically timelike singular static spacetime is quantum mechanically nonsingular when the evolution of any state is uniquely defined for all time. If the evolution is not uniquely defined, then there is some loss of predictability and spacetime is said to be singular in the quantum mechanical sense.

To illustrate this, we consider the propagation of a quantum wave packet satisfying the massive Klein-Gordon wave equation $(\nabla^{\mu}\nabla_{\mu}-\tilde{m}^{2})\psi=0$ in a nonglobally hyperbolic static spacetime. This equation can be written by splitting temporal and spatial parts as

\begin{equation}
\frac{\partial ^{2}\psi }{\partial t^{2}}=VD^{i}\left( VD_{i}\psi\right)-V^{2}\tilde{m}^{2}\psi \label{m1}
\end{equation}
where $V^{2}=-\xi _{\mu }\xi ^{\mu }$ and $D_{i}$ is the spatial covariant derivative on a spatial hypersurface $\Sigma$, which is orthogonal to the static Killing field $\xi^{\mu}$. The spatial operator $A$ is defined on the Hilbert space $\mathcal{H}$ of  square integrable functions on spatial slice $\Sigma$ $(\mathcal{L}^{2}(\Sigma))$,

\begin{equation}
A=-VD^{i}\left(VD_{i}\right)+V^{2}\tilde{m}^{2}.\label{m2}
\end{equation}

It has been shown in \cite{Ashtekar} that, a consistent quantum field theory is possible for a single relativistic particle in a static globally hyperbolic spacetime. This idea has been extended by HM to cover static spacetimes having timelike singularities. In doing so,  the positive frequency solution of  wave operator $A$ with a mass satisfying $0\leq \tilde{m}$ is considered. The wave function for a free relativistic particle is given \textcolor{blue}{in \cite{8}} by

\begin{equation}
i\frac{d\psi }{dt}=\sqrt{A_{E}}\psi , \label{m3}
\end{equation}
whose solution is
\begin{equation}
\psi \left( t\right) =e^{-it\sqrt{A_{E}}}\psi \left( 0\right). \label{m4}
\end{equation}

Here $A_{E}$ denotes the self adjoint extension of the operator defined in Eq. \eqref{m2}. The important point here is to show that  operator $A$ has a unique extension. If the extension is unique, the future time evolution of a wave packet is uniquely defined for all times and this is interpreted as quantum regular. But if the extension is not unique, one must know which extension is used for a time evolution and this in turn will be regarded as quantum singular.

The uniqueness of the extension for the spatial operator $A$ can be checked by considering the solutions of the equation

\begin{equation}
\left( A^{\ast }\pm i\right) \psi =0, \label{m5}
\end{equation}
and it can be shown that there are no square integrable solutions \cite{book1}. This implies that the solution does not belong to Hilbert space and the self-adjoint extension of operator $A$ is unique. Thus, the spatial operator $A$ is essentially self-adjoint.

The spatial part of massive Klein-Gordon equation for the metric given in Eq.(1) is obtained as

\begin{equation}
\begin{aligned}
\emph{A}=&-K^{-4}\Delta^{2\gamma-\gamma^{2}+1}\Sigma^{\gamma^{2}-1} \frac{\partial ^{2}}{\partial r^{2}}-\frac{K^{-4}\Delta^{2\gamma-\gamma^{2}}{\Sigma^{\gamma^{2}-1}} }{r^{2}sin\theta}\frac{\partial }{\partial \theta}\left(sin\theta \frac{\partial }{\partial \theta}\right)-\frac{K^{-4}\Delta^{2\gamma-1}}{r^{2}\sin ^{2}\theta }\frac{\partial ^{2}}{\partial \varphi ^{2}}\\
&-K^{-4}\Delta^{2\gamma-\gamma^{2}}\Sigma^{\gamma^{2}-1}\left(
\frac{2\Delta }{r}+\Delta^{\prime }\left( r\right) \right) \frac{%
\partial }{\partial r}+K^{-2}\Delta^{\gamma}\tilde{m}^{2}. \label{m6}
\end{aligned}
\end{equation}
If we substitute Eq. \eqref{m6} into Eq. \eqref{m5}, we obtain

\begin{equation}
\begin{aligned}
&\left[-K^{-4}\Delta^{2\gamma-\gamma^{2}+1}\Sigma^{\gamma^{2}-1} \frac{\partial ^{2}}{\partial r^{2}}-\frac{K^{-4}\Delta^{2\gamma-\gamma^{2}}{\Sigma^{\gamma^{2}-1}} }{r^{2}sin\theta}\frac{\partial }{\partial \theta}\left(sin\theta \frac{\partial }{\partial \theta}\right)-\frac{K^{-4}\Delta^{2\gamma-1}}{r^{2}\sin ^{2}\theta }\frac{\partial ^{2}}{\partial \varphi ^{2}}\right. \\
&\left.-K^{-4}\Delta^{2\gamma-\gamma^{2}}\Sigma^{\gamma^{2}-1}\left(
\frac{2\Delta }{r}+\Delta^{\prime }\left( r\right) \right) \frac{%
\partial }{\partial r}+K^{-2}\Delta^{\gamma}\tilde{m}^{2}\pm i\right]\psi=0. \label{m7}
\end{aligned}
\end{equation}

Assuming a separable solution in the form of $\psi=f\left( r,\theta\right) e^{\pm ik\varphi}$, in which $k$ is related to the orbital quantum number running for all integers, Eq. \eqref{m7} is transformed into the following form

\begin{equation}
\begin{aligned}
&\left[-K^{-4}\Delta^{2\gamma-\gamma^{2}+1}\Sigma^{\gamma^{2}-1}\frac{\partial^{2} f}{\partial r^{2}}-\frac{K^{-4}\Delta^{2\gamma-\gamma^{2}}{\Sigma^{\gamma^{2}-1}} }{r^{2}sin\theta}\frac{\partial }{\partial \theta}\left(sin\theta \frac{\partial f }{\partial \theta}\right)\pm\frac{K^{-4}\Delta^{2\gamma-1}fk^{2}}{r^{2}\sin ^{2}\theta }\right. \\
&\left.-K^{-4}\Delta^{2\gamma-\gamma^{2}}\Sigma^{\gamma^{2}-1}\left(
\frac{2\Delta }{r}+\Delta^{\prime }\left( r\right) \right) \frac{%
\partial f}{\partial r}+fK^{-2}\Delta^{\gamma}\tilde{m}^{2}\pm if\right]=0. \label{m8}
\end{aligned}
\end{equation}

Our aim is to probe the outermost singularity at $r_{\Delta}=m(1+p)$. When probing this singularity, we will confine the direction of the probe to specific values of $\theta$. Hence, we set $f(r,\theta_{0}=constant)=R(r).$ First, we probe the outermost singularity along the equatorial plane $\theta=\pi/2.$ In this particular case, Eq. \eqref{m8} becomes
\begin{equation}
\begin{aligned}
&(\Delta(r)\Sigma^{-1}(r,\pi/2))^{1-\gamma^{2}}\left[\frac{r^{2}\Delta(r)}{R}\frac{d^{2} R}{dr^{2}}+\frac{r^{2}}{R}\left(
\frac{2\Delta(r) }{r}+\Delta^{\prime }\left( r\right) \right) \frac{%
dR}{dr}\right. \\
&\left.-r^{2}K^{2}(r)\Delta^{\gamma^{2}-\gamma}(r)\Sigma^{1-\gamma^{2}}(r,\pi/2)\tilde{m}^{2}\mp ir^{2}K^{4}(r)\Delta^{\gamma^{2}-2\gamma}(r)\Sigma^{1-\gamma^{2}}(r,\pi/2)\right] \mp k^{2}=0. \label{m9}
\end{aligned}
\end{equation}
In the next stage, the outermost singularity is probed with waves that propagate along the symmetry axis, namely, $\theta=0$ or $\theta=\pi$. In this stage,  Eq. \eqref{m8} becomes ill-defined for all values of $k$ rather than $k=0$ which corresponds to the s-wave mode. Under this restricted condition $(k=0)$, Eq. \eqref{m8} can be written as
\begin{equation}
\begin{aligned}
&\left[\frac{r^{2}\Delta(r)}{R}\frac{d^{2} R}{dr^{2}}+\frac{r^{2}}{R}\left(
\frac{2\Delta(r) }{r}+\Delta^{\prime }\left( r\right) \right) \frac{%
dR}{dr}\right. \\
&\left.-r^{2}K^{2}(r)\Delta^{\gamma^{2}-\gamma}(r)\Sigma^{1-\gamma^{2}}(r,0)\tilde{m}^{2}\mp ir^{2}K^{4}(r)\Delta^{\gamma^{2}-2\gamma}(r)\Sigma^{1-\gamma^{2}}(r,0)\right]=0. \label{m10}
\end{aligned}
\end{equation}

The square integrability of the solutions of Eq. \eqref{m5}  are
checked for each sign by calculating the squared norm, in which the function space on each
$t=$ constant hypersurface $\Sigma _{t}$ is defined as $\mathcal{H}=\left\{
R:\parallel R\parallel <\infty\right\} .$ The general $(n+2)-$ dimensional static spacetime is described by the line element in the form of

\begin{equation}
ds^{2}=-V^{2}dt^{2}+h_{ij}dx^{i}dx^{j}. \label{m11}
\end{equation}
The squared norm for this particular metric can be defined as \cite{9}
\begin{equation}
\Vert R\Vert ^{2}=\frac{q_{0}^{2}}{2}\int_{\Sigma _{t}}V^{-1}RR^{\ast }d^{n+1}x\sqrt{h}, \label{mm}
\end{equation}%
in which $q_{0}^{2}$ is a positive constant and $h_{ij}$ is the spatial part of the line element. Here, $h$ is the determinant of the spatial part. The spatial operator $A$ has a unique self-adjoint extension if neither of the
solutions of Eq. \eqref{m5} are square integrable over all space. \\

At this stage, it is important to emphasize the choice of the function space to be used in the analysis. Our choice is the usual square integrable $L^{2}$ Hilbert space of quantum mechanics. Another alternative is the first Sobolev space $H^{1}$, proposed by Ishibashi and Hosoya \cite{9}. However, Sobolev space is not the usual quantum – mechanical Hilbert space. The main difference is in the definition of the norm. In the Sobolev space, the norm including both the wave function and its derivative need to be square integrable. If the function space is the natural linear function space of quantum mechanics, which is the one used in this study, the norm calculation involves the wave function only. This in turn brings a strong condition on the self-adjointness of the spatial part of the Hamiltonian operator $A$. The use of Sobolev space, however, weakens this strong condition with the additional term that involves derivative. As a result,  while the wave function is square integrable, its derivative may not be square integrable. For example, the negative mass Schwarzchild solution and $5-dimensional$ string solution studied in \cite{9} (which were studied in \cite{8}, and found as quantum singular) are shown to be quantum regular. Similar results are also obtained in \cite{20}, for the analysis of naked singularities in $2+1 - dimensional$ $BTZ$ metrics coupled with linear electrodynamics and dilaton fields.

In the following subsections, the quantum nature of the non-trivial naked singularities developed in the charged and uncharged ZV - metrics will be investigated. The outermost singularity, which is valid for all values of $\gamma$, except for zero and one, will be investigated on the equatorial plane. This analysis will be followed by the investigation of the quantum nature of the directional singularities developed at the poles.

\subsection{Quantum singularity analysis of the outermost singularity}

Our aim in this subsection is to investigate the outermost singularity located at $r_{\Delta}=m(1+p)$ when $\theta=\pi/2$, both for the charged and uncharged ZV metrics.
\subsubsection{For the charged $ZV$ solution:}

When we substitute $\theta=\pi/2$ into Eq.(3), Eq. \eqref{m9} can be written as

\begin{equation}
R^{\prime \prime }+\frac{\left( r^{2}\Delta\right) ^{\prime }}{\Delta r^{2}}R^{\prime }-%
\Delta^{\gamma^{2}-1}\sigma^{1-\gamma ^{2}}\left[ \pm\frac{k^{2} }{\Delta r^{2}}+K^{2}\Delta ^{-\gamma}\tilde{m}^{2}\pm iK^{4}\Delta^{-2\gamma}%
\right] R=0, \label{m12}
\end{equation}

where $\Sigma(r,\pi/2)=\sigma(r)=1-\frac{2m}{r}+\frac{m^{2}}{r^{2}}$. The solution of Eq. \eqref{m12} will be investigated for two different limiting cases, $r\rightarrow \infty $ and $r\rightarrow m(1+p)$ (near the singularity).

In the case when $r\rightarrow \infty$, the metric functions behave as

\begin{equation}
\begin{aligned}
&\sigma(r)=\Delta(r)\approx 1-\frac{2m}{r},\\
&K\approx 2p. \label{m13}
\end{aligned}
\end{equation}
In this limit, Eq. \eqref{m12} simplifies to
\begin{equation}
R^{\prime \prime }+\frac{2}{r}R^{\prime }+\left( (2p)^{2}\tilde{m}^{2}\pm (2p)^{4}i\right)
R=0, \label{m14}
\end{equation}%
whose solution is given by%
\begin{equation}
R(r)=\frac{C_{1}}{r}\sin \kappa r+\frac{C_{2}}{r}\cos \kappa r. \label{m15}
\end{equation}%
Here, $\kappa =\sqrt{(2p)^{2}\tilde{m}^{2}\pm (2p)^{4}i}$ and $C_{1},$ $C_{2}$ are integration constants. Substituting Eq. \eqref{m15}  into Eq. \eqref{mm} leads to
\begin{equation}
\Vert R\Vert ^{2}\sim \int_{const}^{\infty }\frac{(RR^{*}) dr}{\left(1- \frac{2m}{r}\right)^{2\gamma-1}}
.\label{m16}
\end{equation}%
Note that for the practical purposes we choose $C_{1}=C_{2}=1$. In order to evaluate this integral, we first expand the denominator by using the binomial expansion for large values of $r$,  taking into account  the dominant terms only.  Since $sin(\kappa r)$ and $cos(\kappa r)$ are complex valued trigonometric functions, we transform $RR^{*}$ to the following form with the help of $\sqrt{a\textcolor{blue}{\pm} bi}=\pm \left( \sqrt{\frac{\vert z \vert+a}{2}}\textcolor{blue}{\pm}  i\sqrt{\frac{\vert z \vert-a}{2}} \right)$, where $\vert z \vert=\sqrt{a^{2}+b^{2}}$ and $\textcolor{blue}{b>0}$,

\begin{equation}
\Vert R\Vert ^{2}\sim \int_{const}^{\infty }\frac{rcosh(2\bar{\kappa}_{1}r)}{r-2m(2\gamma-1)}dr+\int_{const}^{\infty }\left( \frac{r\sin(2 \bar{\kappa}_{2} r)}{%
r-2m(2\gamma-1)}\right) dr, \label{m17}
\end{equation}

in which $\bar{\kappa}_{1}=\pm \sqrt{2p\sqrt{\tilde{m}^{2}+(2p)^{4}}-2p\tilde{m}^{2}}$ and $\bar{\kappa}_{2}=\pm \sqrt{2p\sqrt{\tilde{m}^{2}+(2p)^{4}}+2p\tilde{m}^{2}}$. \\
The calculation of these integrals are shown in Appendix A. \\
It is worth to state that the calculations carried out for the large $r$ limit in the present study is just for the sake of completness of the requirement of the analysis for all space ranging from outermost singularity to the infinity. Since the $ZV$ metric is asymptotically flat it is not necessary to investigate the behavior at large values of $r$. As a result, determining the essential self adjointness of the operator $A$ simply requires assessing the square integrability of the solutions in the vicinity of the outermost singularity. The following calculations illustrates this case in detail. \\

In the case when $ r \rightarrow r_{\Delta}=m(1+p)$, Eq. \eqref{m12} transforms to

\begin{equation}
\frac{d^{2}R}{dr^{2}}+\frac{1}{r-r_{+}}\frac{dR}{dr}-\frac{p^{2}}{(1+p)^{2}}\left[\frac{\pm k^{2}\Delta^{\gamma^{2}-2}}{r^{2}} +K^{2}\tilde{m}^{2}\Delta^{\gamma^{2}-\gamma-1}\pm i K^{4}\Delta^{\gamma^{2}-2\gamma-1}\right]R=0. \label{m18}
\end{equation}

This equation will be splitted into three independent equations by considering the rate of divergence in connection with the deformation parameter $\gamma$. Thus, we have

\begin{equation}
\frac{d^{2}R}{dr^{2}}+\frac{1}{r-r_{\Delta}}\frac{dR}{dr}+iH_{\gamma}(r)R=0, \label{m19}
\end{equation}
in which

\begin{equation}
H_{\gamma}(r)=
\begin{cases}
\frac{\beta_{1}}{(r-r_{\Delta})^{2-\gamma^{2}}}&,0<\gamma<1/2 \\
\frac{\beta_{2}}{(r-r_{\Delta})^{1+2\gamma-\gamma^{2}}}&,\gamma>1/2 \: \: \: \: \: \: \: \: \:\: \: \: \: \: \: \: \: \: \: \: \: \: \: \: \: \: \: \: \: \: \: \: \\
\frac{\beta_{3}}{(r-r_{\Delta})^{7/4}}&,
\gamma=1/2 \: \: \: \: \: \: \: \: \:\: \: \: \: \: \: \: \: \: \: \: \: \: \: \: \: \: \: \: \: \: \: \: \label{m20}
\end{cases},
\end{equation}
where

\begin{equation}
\begin{aligned}
\beta_{1}=\frac{\pm ip^{2}k^{2}(2mp)^{\gamma^{2}-2}}{(1+p)^{2}r_{\Delta}^{2\gamma^{2}-2}}\\
\beta_{2}=\pm\left(\frac{2p}{1+p} \right)^{4\gamma}\frac{p^{2}(1+p)^{2}(2mp)^{\gamma^{2}-2\gamma-1}}{r_{\Delta}^{2\gamma^{2}-4\gamma-2}}\\
\beta_{3}=\frac{p^{2}r_{\Delta}^{7/4}}{(2mp)^{7/4}}\left[ (1+p)^{4}\left(\frac{2p}{1+p} \right)^{4\gamma}\mp\frac{k^{2}}{m^{2}(1+p)^{2})} \right]. \label{m21}
\end{aligned}
\end{equation}

The solution for each interval of $\gamma$ is given by

\begin{equation}
R(r) = \begin{cases}
a_{1}K_{0}(\eta_{1}(r-r_{\Delta})^{\gamma^{2}/2})+(a_{2}){}_{0}F_{1}(;1;\eta_{2}(r-r_{\Delta})^{\gamma^{2}})&,0<\gamma<1/2 \\
a_{3}K_{0}(\eta_{3}(r-r_{\Delta})^{\frac{\gamma^{2}-2\gamma+1}{2}})+(a_{4}){}_{0}F_{1}(;1;\eta_{4}(r-r_{\Delta})^{\gamma^{2}-2\gamma+1})&,\gamma>1/2 \: \: \: \: \: \: \: \: \:\: \: \: \: \: \: \: \: \: \: \: \: \: \: \: \: \: \: \: \: \: \: \: \\
a_{5}K_{0}(\eta_{5}(r-r_{\Delta})^{1/8})+(a_{6}){}_{0}F_{1}(;1;\eta_{6}(r-r_{\Delta})^{1/4})&,\gamma=1/2 \: \: \: \: \: \: \: \: \:\: \: \: \: \: \: \: \: \: \: \: \: \: \: \: \: \: \: \: \: \: \: \: \label{m22}
\end{cases}
\end{equation}
in which $a_{1}$ to $a_{6}$ are integration constants, $K_{0}$ is the first kind modified Bessel function, ${}_{0}F_{1}$ is the confluent hypergeometric function, $\eta_{1}=\frac{2(-1)^{7/4}\sqrt{\beta_{1}}}{\gamma^{2}}$, $\eta_{2}=\frac{-i\beta_{1}}{\gamma^{4}}$, $\eta_{3}=\frac{2(-1)^{3/4}\sqrt{\beta_{2}}}{2\gamma-1-\gamma^{2}}$, $\eta_{4}=\frac{-i\beta_{2}}{(2\gamma-1-\gamma^{2})^{2}}$, $\eta_{5}=8\sqrt{\beta_{3}}$ and $\eta_{6}=-16\beta_{3}$.   \\
The square norm for each case can be written as

\begin{equation}
\Vert R\Vert ^{2}\sim \begin{cases}
\left(\frac{2p}{1+p} \right)^{4\gamma}\frac{p^{2-2\gamma^{2}}(1+p)^{2\gamma^{2}-1}}{(2mp)^{2\gamma-\gamma^{2}}r_{\Delta}^{2\gamma^{2}-4\gamma-2}}\int_{const}^{r_{\Delta} }\frac{RR^{*}dr}{(r-r_{\Delta})^{2\gamma-\gamma^{2}}}&,0<\gamma<1/2  \\
\left(\frac{2p}{1+p} \right)^{4\gamma}\frac{p^{2-2\gamma^{2}}(1+p)^{2\gamma^{2}-1}}{(2mp)^{-2\gamma+\gamma^{2}}r_{\Delta}^{2\gamma^{2}-4\gamma-2}}\int_{const}^{r_{\Delta} }(r-r_{\Delta})^{\gamma^{2}-2\gamma}RR^{*}dr&,\gamma>1/2 \\
\left(\frac{2p}{1+p} \right)^{2}\frac{p^{3/2}(1+p)^{-1/2}r_{\Delta}^{7/2}}{(2mp)^{3/4}}\int_{const}^{r_{\Delta} }\frac{RR^{*}dr}{(r-r_{\Delta})^{3/4}}&,\gamma=1/2 \: \: \: \: \: \: \: \: \:\: \: \: \: \: \: \: \: \: \: \: \: \: \: \: \: \: \: \: \: \: \: \: \label{m23}
\end{cases}
\end{equation}

Detailed calculations of Eq. \eqref{m23} concerning Cauchy product are given in Appendix B. The integral calculations for each range of the parameter $\gamma$ revealed that the solutions are square integrable and belong to the Hilbert space. As a consequence, the outermost singularity becomes quantum mechanically singular for all values of  $\gamma$.

\subsubsection{The uncharged $ZV$ solution:}

If we choose $q=0$ and consider the scaling factor to be $t \rightarrow 2t$ on the time coordinate, Eq. \eqref{m12} can be written as

\begin{equation}
R^{\prime \prime }+\frac{\left( r^{2}\Delta_{zv}\right) ^{\prime }}{\Delta_{zv} r^{2}}R^{\prime }-%
\Delta_{zv}^{\gamma^{2}-1}\sigma_{zv}^{1-\gamma ^{2}}\left[ \frac{\pm k^{2}}{\Delta_{zv} r^{2}}+\Delta_{zv} ^{-\gamma}\tilde{m}^{2}\pm i\Delta_{zv}^{-2\gamma}%
\right] R=0. \label{m24}
\end{equation}
Here, $\Delta_{zv}=1-\frac{2m}{r}$ and $\sigma_{zv}=1-\frac{2m}{r}+\frac{m^{2}}{r^{2}}$. The  solution of Eq. \eqref{m24} will be analysed for $r\rightarrow \infty $ and $r\rightarrow 2m$ (near singularity), respectively.

In the case when $ r \rightarrow \infty,$ using the approximate metric functions given in Eq. \eqref{m13}, Eq. \eqref{m24} simplifies to

\begin{equation}
R^{\prime \prime }+\frac{2}{r}R^{\prime }+\left(\tilde{m}^{2}\pm i\right)
R=0. \label{m25}
\end{equation}%
The solutions of Eq. \eqref{m25} can be written as
\begin{equation}
R(r)=\frac{C_{5}}{r}\sin \nu r+\frac{C_{6}}{r}\cos \nu r, \label{m26}
\end{equation}%
where $\nu =\sqrt{\tilde{m}^{2}\pm i}$ and $C_{5},$ $C_{6}$ are integration constants. Since the obtained solution is similar to the solution found earlier in Eq. \eqref{m15}, our focus will be on the vicinity of the singularity. 

In the case when $ r \rightarrow 2m$, we introduce a new variable $x$ defined by   $ x \equiv r-2m \rightarrow 0$. Then, one can express the metric functions in terms of the new variable as

\begin{equation}
\begin{aligned}
&\sigma_{zv}(x)=\frac{x}{x+2m}+\frac{m^{2}}{\left( x+2m\right)^{2}}, \\
&\Delta_{zv}(x)= \frac{x}{x+2m}. \label{m27}
\end{aligned}
\end{equation}
If we substitute metric functions presented above into the differential Eq. \eqref{m24} and consider the dominant terms  by taking into account that the parameter $\gamma$ change in certain intervals, the differential Eq. \eqref{m24} becomes

\begin{equation}
R^{\prime \prime }+\frac{1}{x}R^{\prime }+ax^{\mu}R=0, \label{m28}
\end{equation}%
in which

\begin{equation}
ax^{\mu} = \begin{cases}
\pm i (4m^{2})^{\gamma}\left(\frac{2}{m} \right)^{\gamma^{2}-1}x^{\gamma^{2}-2\gamma-1}&,\frac{1}{2}<\gamma<\infty  \;\; and \;\; \gamma \neq 1 \\
 \left(\frac{2}{m} \right)^{\gamma^{2}-1}\left(\frac{k^{2}}{2m} \right)x^{\gamma^{2}-2}&,0<\gamma<\frac{1}{2} \: \: \: \: \: \: \: \: \:\: \: \: \: \: \: \: \: \: \: \: \: \: \: \: \: \: \: \: \: \: \: \: \\
\left(\frac{2}{m} \right)^{\gamma^{2}-1}\left( \pm i (4m^{2})^{\gamma}\pm\frac{k^{2}}{2m} \right)x^{-7/2}
&,\gamma=\frac{1}{2} \: \: \: \: \: \: \: \: \:\: \: \: \: \: \: \: \: \: \: \: \: \: \: \: \: \: \: \: \: \: \: \: \label{m29}
\end{cases}
\end{equation}

The general solution of Eq. \eqref{m28}  is found by

\begin{equation}
R(x)=C_{7}J_{0}\left( \frac{2\sqrt{a}sign(\mu+1)}{\mu+2}x^{\frac{\mu+2}{2}}\right) +C_{8}Y_{0}\left(\frac{2\sqrt{a}sign(\mu+1)}{\mu+2}x^{\frac{\mu+2}{2}}\right). \label{m30}
\end{equation}%
Note that $C_{7}$ and  $C_{8}$\ are integration constants, $J$ and $Y$ are respectively the Bessel functions of the first and second kinds,  with the signum function $sing(\mu+1)$. The behaviour of the Bessel functions for real $\nu \geq 0$ as $ x \rightarrow 0 $ can be stated as \cite{book2}

\begin{equation}
\begin{aligned}
J_{\nu}(x) &\sim \frac{1}{\Gamma(\nu +1)}\left ( \frac{x}{2}\right )^{\nu}\\
Y_{\nu}(x) &\sim \begin{cases}
\frac{2}{\pi}\left[ ln \left ( \frac{x}{2} \right ) +\tilde{\gamma} \right ]&,\nu =0 \;and \; \tilde{\gamma} \cong 0.5772\\
-\frac{\Gamma(\nu)}{\pi}\left(\frac{2}{x}\right)^{\nu}&,\nu\neq 0 \: \: \: \: \: \: \: \: \:\: \: \: \: \: \: \: \: \: \: \: \: \: \: \: \: \: \: \: \: \: \: \: \label{m31}
\end{cases}
\end{aligned}
\end{equation}
Hence, the solution  becomes

\begin{equation}
R(x) \sim \bar{C_{1}}+\bar{C_{2}}ln(x) \label{m32}
\end{equation}
with $\bar{C_{1}}=\frac{C_{7}}{\Gamma(1)}+\frac{2\tilde{\gamma}C_{8}}{\pi}+\frac{2C_{8}}{\pi}ln\left(\frac{\sqrt{a}sing(\mu+1)}{\mu+2} \right)$ and $\bar{C_{2}}=\frac{C_{8}}{\pi}(\mu+2)$ . Once Eq. \eqref{m32} is substituted into the squared norm \eqref{mm},  it becomes

\begin{equation}
\begin{aligned}
\| R \|^{2}& \sim \left( m\right)^{2\gamma-\gamma^{2}+2}\left(2 \right)^{\gamma^{2}+2\gamma} \int_{const.}^{0}x^{\gamma^{2}-2\gamma}|\bar{C_{1}}+\bar{C_{2}}ln(x)|^{2}dx \\
&\sim\left( m\right)^{2\gamma-\gamma^{2}+2}\left\{ A\int_{const.}^{0}x^{\gamma^{2}-2\gamma}dx+B\int_{const.}^{0}x^{\gamma^{2}-2\gamma}ln(x)dx+C\int_{const.}^{0}x^{\gamma^{2}-2\gamma}ln^{2}(x)dx \right\} \label{m33}
\end{aligned}
\end{equation}
where $A=\left(2 \right)^{\gamma^{2}+2\gamma} \bar{C_{1}}^{2}$, $B=\left(2 \right)^{\gamma^{2}+2\gamma+1} \bar{C_{1}}\bar{C_{2}}$ and $C=\left(2 \right)^{\gamma^{2}+2\gamma} \bar{C_{2}}^{2}$. The spatial part of the Hamiltonian operator $A$ is not essentially self-adjoint due to the following reasons. $i)$ The results of the integrals are proportional to $x^{a}ln^{b}(x)$ terms. $ii)$ $lim_{x\rightarrow 0}x^{a}ln^{b}(x)$ is convergent for $a>0$. $iii)$ The square norm convergent as $\gamma\neq1$.

The quantum singularity analysis of the outermost singularity developed both for the charged and uncharged ZV solutions reveals that irrespective of the value of $\gamma$ except for $\gamma=0$ and $\gamma=1$, the classically singular character continues to remain singular even if it is probed with quantum wave packets obeying the Klein-Gordon equations.

\subsection{Quantum singularity analysis of directional singularities}

Directional singularities are known to be characteristic properties of the $\gamma$ - metrics, and this behavior is also valid for the charged version of the ZV - metrics. From a classical point of view, the spacetime appears to be regular for the deformation parameter $\gamma\geq2$, if we first let $\theta=0$ (north pole), or $\theta=\pi$ (south pole), and then approach to the outermost singularity. However, the spacetime becomes singular whenever the deviation parameter is in the interval of $0<\gamma<2$, but not equal to $1$. On the other hand if we  first let $r_{\Delta}=m(1+p)$ and then take the limits $\theta=0$  or $\theta=\pi$ , the spacetime is singular whenever $\gamma>0$, but not equal to $1$.
In this subsection, our aim is to investigate classical directional singularities within the context of quantum mechanics. As in the previous subsections, directional singularities will be probed with quantum wave packets obeying the Klein-Gordon equation.

\subsubsection{For the charged $ZV$ solution:}

As we clarified earlier, the directional singularities on the symmetry axis will be probed with s-waves only. In this particular case that corresponds to $k=0$, Eq. \eqref{m10} is transformed to

\begin{equation}
R^{\prime \prime }+\frac{\left( r^{2}\Delta\right) ^{\prime }}{\Delta r^{2}}R^{\prime }-K^{2}\left[\Delta ^{-\gamma}\tilde{m}^{2}\pm iK^{2}\Delta^{-2\gamma} \label{m34}
\right] R=0.
\end{equation}

We first consider the case when $r \rightarrow \infty$. For this scenario, the approximate metric functions defined in Eq. \eqref{m13} are used in Eq. \eqref{m34}, which leads to

\begin{equation}
R^{\prime \prime }+\frac{2}{r}R^{\prime }+\left( (2p)^{2}\tilde{m}^{2}\pm (2p)^{4}i\right)
R=0. \label{m35}
\end{equation}%
Since Eq. \eqref{m35} coincides with Eq. \eqref{m15}, which has previously been studied.

In the case when $ r \rightarrow r_{\Delta}=m(1+p)$, by considering the dominant terms with respect to the value of $\gamma$,  Eq. \eqref{m34} becomes

\begin{equation}
\frac{d^{2}R}{dr^{2}}+\frac{1}{r-r_{\Delta}}\frac{dR}{dr}+U_{\gamma}(r)R=0, \label{m36}
\end{equation}
in which

\begin{equation}
U_{\gamma}(r) =\frac{ib_{2}}{(r-r_{\Delta})^{2\gamma}}, \label{m37}
\end{equation}
where $b_{2}=\pm(1+p)^{4}\left(\frac{2p}{1+p} \right)^{4\gamma}\frac{ r_{\Delta}^{4\gamma}}{(2mp)^{2\gamma}}$.

The solution of Eq. \eqref{m36} is given by

\begin{equation}
R(r) = d_{3}K_{0}(\nu_{1}(r-r_{\Delta})^{1-\gamma})+(d_{4}){}_{0}F_{1}(;1;\nu_{2}(r-r_{\Delta})^{2-2\gamma}) \label{m37}
\end{equation}
in which $\nu_{1}=\frac{\sqrt{b_{2}}(-1)^{3/4}}{\gamma-1}$ and $\nu_{2}=\frac{-ib_{2}}{4(\gamma-1)^{2}}$. The square norm of this case can be written as

\begin{equation}
\Vert R\Vert ^{2}\sim \left(\frac{2p}{1+p} \right)^{4\gamma}(1+p)^{4}(2mp)^{2\gamma-1}r_{\Delta}^{4\gamma}\int_{const}^{r_{\Delta} }\frac{RR^{*}dr}{(r-r_{\Delta})^{2\gamma-1}} \label{m38}
\end{equation}
in which $d_{3}$ and $d_{4}$ are taken to be one. The integrals are calculated by changing the variable of integration $u=r-r_{\Delta}$, followed by considering that $u$ is very small in the vicinity of $r\rightarrow r_{\Delta}$. The Cauchy product law is then used as in the previous calculations and the square norm equation becomes

\begin{equation}
\begin{aligned}
\Vert R\Vert ^{2} \sim &\bar{\alpha_{2}}\sum_{k=0}^{\infty}c_{k}\int_{const}^{0 }\left(ln^{2}\frac{\nu_{1}u^{1-\gamma}}{2} \right)\left(\frac{\nu_{1}u^{1-\gamma}}{2} \right)^{k}\frac{du}{u^{2\gamma-1}}\\
&+ \bar{\alpha_{2}}\sum_{l=0}^{\infty}c_{l}\int_{const}^{0 }\left(\nu_{2}u^{2-2\gamma} \right)^{l}\frac{du}{u^{2\gamma-1}}\\
&-2\bar{\alpha_{2}}\sum_{n=0}^{\infty}c_{n}\int_{const}^{0 }\left(ln\frac{\nu_{1}u^{1-\gamma}}{2} \right)\left(u^{2-2\gamma} \right)^{n}\frac{du}{u^{2\gamma-1}} \label{m39}
\end{aligned}
\end{equation}
where $\bar{\alpha_{2}}=\left(\frac{2p}{1+p} \right)^{4\gamma}(1+p)^{4}(2mp)^{2\gamma-1}r_{\Delta}^{4\gamma}$.
Since the power of $u$ is important for the analysis, we split the deformation parameter $\gamma$ into two subcases; namely, $0<\gamma<1$ and $\gamma>1$. Note that, other parts containing the $ln$ function are again convergent as shown in the previous section. \\
First we consider the case of $0<\gamma<1$. In this case,  the power of $u$, which is $(1-\gamma)>0$ is always positive and since $u$ is positive and very small, we can define an inequality  $u^{2l(1-\gamma)-(2\gamma-1)}\leq u^{-(2\gamma-1)}$. Since the integral $\int_{const}^{0 }u^{1-2\gamma}du=\frac{u^{2-2\gamma}}{2-2\gamma}\left\vert _{const}^{0 }\right< \infty $ is convergent, $\int_{const}^{0 } u^{2l(1-\gamma)-(2\gamma-1)}du$ is also convergent as a requirement of the comparison test. Hence, the directional singularities along the axis become quantum singular when the deformation parameter is in the interval of $0<\gamma<1$.

Now, let us consider the case when $\gamma>1$. In this case,  $(1-\gamma)<0$ and the inequality for the comparison test can be defined as $u^{-(2\gamma-1)}\leq u^{2l(1-\gamma)-(2\gamma-1)}$. A careful analysis reveals that when $\gamma>1$, the integral $\int_{const}^{0 }u^{1-2\gamma}du=\frac{u^{2-2\gamma}}{2-2\gamma}\vert _{const}^{0 } \rightarrow \infty $ fails to be square integrable. As a result of this analysis, the directional singularity, when approached from the north pole ($\theta=0$), becomes quantum mechanically regular as long as $\gamma>1$. If we compare this with the classical analysis ($\gamma\geq2$), there is a partial healing on the singularities.

\subsubsection{For the uncharged $ZV$ solution:}

Eq. \eqref{m10} when $q=0$ is

\begin{equation}
R^{\prime \prime }+\frac{\left( r^{2}\Delta_{zv}\right) ^{\prime }}{\Delta_{zv} r^{2}}R^{\prime }-\left[ \Delta_{zv} ^{-\gamma}\tilde{m}^{2}\pm i\Delta_{zv}^{-2\gamma}%
\right] R=0. \label{m40}
\end{equation}

In the case when $ r \rightarrow \infty,$ Eq. \eqref{m40} becomes

\begin{equation}
R^{\prime \prime }+\frac{2}{r}R^{\prime }+\left(\tilde{m}^{2}\pm i\right)
R=0. \label{m41}
\end{equation}%

This equation is exactly the same as Eq. \eqref{m15}.

Now, let us investigate the case when $ r \rightarrow 2m \Leftrightarrow x \equiv r-2m \rightarrow 0$ .  If we use the metric function defined in Eq. \eqref{m27} in Eq. \eqref{m40}, we have

\begin{equation}
R^{\prime \prime }+\frac{1}{x}R^{\prime }+bx^{-2\gamma}R=0, \label{m42}
\end{equation}%
in which $b=\pm i(4m^{2})^{\gamma}$. The solution of Eq. \eqref{m42} is given by

\begin{equation}
R(x)=q_{3}J_{0}\left( \frac{2\sqrt{b}sign(1-2\gamma)}{2-2\gamma}x^{1-\gamma}\right) +q_{4}Y_{0}\left(\frac{2\sqrt{b}sign(1-2\gamma)}{2-2\gamma}x^{1-\gamma}\right), \label{m43}
\end{equation}%
where $q_{3}$ and  $q_{4}$ are integration constants. When we use the asymptotic approximations of the Bessel functions defined in Eq. \eqref{m31}, Eq. \eqref{m43} reduces to

\begin{equation}
R(x) \sim \bar{q_{3}}+\bar{q_{4}}ln(x) \label{m44}
\end{equation}
with $\bar{q_{3}}=\frac{q_{3}}{\Gamma(1)}+\frac{2\tilde{\gamma}q_{4}}{\pi}+\frac{2q_{4}}{\pi}ln\left(\frac{\sqrt{b}sing(1-2\gamma)}{2-2\gamma} \right)$ and $\bar{q_{4}}=\frac{q_{4}}{\pi}(2-2\gamma)$ . The squared norm for solution \eqref{m44} is given by

\begin{equation}
\begin{aligned}
\| R \|^{2}& \sim \left( 2m\right)^{2\gamma+1} \int_{const.}^{0}x^{1-2\gamma}|\bar{q_{3}}+\bar{q_{3}}ln(x)|^{2}dx \\
&\sim\bar{A}\int_{const.}^{0}x^{1-2\gamma}dx+\bar{B}\int_{const.}^{0}x^{1-2\gamma}ln(x)dx+\bar{C}\int_{const.}^{0}x^{1-2\gamma}ln^{2}(x)dx \label{m45}
\end{aligned}
\end{equation}
where $\bar{A}=\left( 2m\right)^{2\gamma+1} \bar{q_{3}}^{2}$, $\bar{B}=\left( 2m\right)^{2\gamma+1} \bar{q_{3}}\bar{q_{4}}$ and $\bar{C}=\left( 2m\right)^{2\gamma+1}\bar{q_{4}}^{2}$. Since the results of the last two integrals are proportional to  $x^{a}ln^{b}(x)$ for $b=1,2$ terms, the $lim_{x\rightarrow 0}x^{a}ln^{b}(x)$ is finite , and also because $\gamma\neq 1$, the square norm converges. Therefore, the spacetime is quantum mechanically singular. However, the square integrability analysis has revealed that whenever $\gamma>1$, the squared norm diverges according to the first integral. This implies that the solution for this particular choice fails to be square integrable. On the other hand, if $\gamma<1$, then the solution is square integrable.

As a result of this analysis, classically singular sector in $1<\gamma<2$ of the directional singularity of the uncharged ZV metric becomes quantum mechanically regular.

\section{Discussion}\label{4}

In this paper, we have investigated the formation of timelike naked singularities developed in the charged and uncharged $ZV$ spacetimes. Our analysis is based on the Horowitz and Marolf definition of quantum singularities adopted for static spacetimes which utilizes quantum wave packets obeying the Klein-Gordon equation. \\
In doing so, we have focused on two particular types of classical singularity. One of them is the singularity that is located on the equatorial plane at  $r_{\Delta}=m(1+p)$, which is abbreviated as the outermost singularity, valid for all values of the deformation parameter $\gamma$ except zero and one. The other one is the directional singularity, peculiar to  $\gamma - metrics$, which becomes regular if the singular point is approached along the north or south pole. The regularity of this directional singularity crucially depends on the deformation parameter $\gamma\geq2$.

In our analysis, we have used Klein-Gordon quantum wave packets for probing the singularities. It has been shown by rigorous calculations that the outermost singularity remains quantum mechanically singular for all values of  $\gamma$, regardless of the non-spherical compact object being charged or uncharged. However, precise and detailed analysis of the poles  associated with the directional singularities exhibits "partial healing" such that the range of the regularity condition on the deformation parameter is extended to $\gamma>1$. To be more clear, the classically  singular sector becomes quantum mechanically regular both for charged and uncharged $ZV$ metrics when $1<\gamma<2$. However, it is important to emphasize that the "partial healing" is possible if the analysis is restricted  to the s-wave mode only. In the generic case, since  Eq. \eqref{m8} is ill-defined for $k\neq0$, the classical directional singularities along the symmetry axis remains quantum mechanically singular. \\

As a final remark, as long as the real compact objects are concerned, understanding the charged or uncharged $ZV$ metrics that describe the exterior geometry of non-spherical compact objects is very important. But, the most important thing is to understand the singularity structure. Is the "partial healing" on the directional singularity  peculiar to oblate objects for the s-wave mode only ? What will happen if spinorial probes are used ? It would be interesting to further investigate the singularity structure of the $ZV$ spacetimes in view of these findings.\\

\textbf{Appendix A. Detailed Calculation for Norm Squared at $r\rightarrow\infty$ in the Charged $ZV$ Spacetime}\\

We can use the comparison test to analyze the convergence character of the first integral in Eq. \eqref{m17}. The series expansion of the hyperbolic function is as follows.

\begin{equation}
\cosh \left( 2\bar{\kappa}_{1} r\right) =\sum_{n=0}^{\infty }\frac{%
\left( 2\bar{\kappa}_{1} r\right) ^{2n}}{\left( 2n\right) !}.
\end{equation}%

The first integral becomes

\begin{equation}
\begin{aligned}
\int_{const}^{\infty }\frac{rcosh(2\bar{\kappa}_{1}r)}{r-2m(2\gamma-1)}dr&=\int_{const}^{\infty }\left( \frac{r}{r-2m(2\gamma-1)}\right) \left\{
\sum_{n=0}^{\infty }\frac{%
\left( 2\bar{\kappa}_{1} r\right) ^{2n}}{\left( 2n\right) !}\right\} dr\\
&=\sum_{n=0}^{\infty }\frac{%
\left( 2\bar{\kappa}_{1} \right) ^{2n}}{\left( 2n\right) !}%
\int_{const}^{\infty }\left( \frac{r^{b}}{r-2m(2\gamma-1)}\right) dr,
\end{aligned}
\end{equation}%

where $b=2n+1$. The convergence character of the last integral can be analyzed by using the comparison test. For this purpose, the following inequality is studied

\begin{equation}
0\leq \frac{t+2m(2\gamma-1)}{t}\leq \frac{\left( t+2m(2\gamma-1)\right) ^{b}}{t},
\end{equation}%
where $t=r-2m(2\gamma-1)$. The integral of $\frac{t+2m(2\gamma-1)}{t}$ is calculated as
\begin{equation}
\int_{c}^{\infty }\left( \frac{t+2m(2\gamma-1)}{t}\right) dt=\left( t+2m(2\gamma-1)\ln \left\vert
t\right\vert \right) \left\vert _{const}^{\infty }\right. \rightarrow \infty.
\end{equation}%

According to the comparison test, the divergence of the integral $%
\int_{c}^{\infty }\left( \frac{t+2m(2\gamma-1)}{t}\right) dt$ implies the divergence of
$\int_{c}^{\infty }\frac{\left( t+2m(2\gamma-1)\right) ^{b}}{t}dt$ .
Finally, let us analyze the convergence character of the series in front of the integral with the help of the ratio test. If we construct the expression $\rho=lim_{n\rightarrow \infty}\vert\frac{ a_{n+1}}{a_{n}}\vert=lim_{n\rightarrow \infty}\frac{(2\bar{\kappa}_{1})^{2}}{2n+1}=0$ for the convergence analysis, we see that for $0=\rho<1$, this limit converges and hence the series converges according to the ratio test. \\
The second integral is evaluated by using the comparison test, especially
developed for the improper integrals. We replace $\sin \left( 2 \bar{\kappa}_{2} r\right) $ with its power series expansion,%

\begin{equation}
\sin \left( 2\bar{\kappa}_{2} r\right) =\sum_{n=0}^{\infty }\left( -1\right) ^{n}sign(\bar{\kappa}_{2})\frac{%
\left( 2\omega_{2} r\right) ^{2n+1}}{\left( 2n+1\right) !},
\end{equation}%
where $\omega_{2}=\sqrt{2p\sqrt{\tilde{m}^{2}+(2p)^{4}}+2p\tilde{m}^{2}}$ and $sign$ is the signum function.
The second integral becomes%

\begin{equation}
\begin{aligned}
I&=\int_{const}^{\infty }\left( \frac{r}{r-2m(2\gamma-1)}\right) \left\{
\sum_{n=0}^{\infty }\left( -1\right) ^{n}sign(\bar{\kappa}_{2})\frac{\left( 2\omega_{2} r\right)
^{2n+1}}{\left( 2n+1\right) !}\right\} dr\\
&=\sum_{n=0}^{\infty }\left(
-1\right) ^{n}sign(\bar{\kappa}_{2})\frac{\left( 2\omega_{2} \right) ^{2n+1}}{\left( 2n+1\right) !}%
\int_{const}^{\infty }\left( \frac{r^{a}}{r-2m(2\gamma-1)}\right) dr,
\end{aligned}
\end{equation}%
in which $a=2n+2$. The result of the integral comparison test similar to the calculations made above is divergent, and also, since $\rho=lim_{n\rightarrow \infty}\vert\frac{ a_{n+1}}{a_{n}}\vert=lim_{n\rightarrow \infty}\frac{(2\omega_{1})^{2}}{(2n+3)(2n+2)}=0<1$, it is convergent according to the ratio test.\\

\textbf{Appendix B. Detailed Calculation for Norm Squared at $r\rightarrow r_{\Delta}=m(1+p)$ in the Charged $ZV$ Spacetime}\\

In order to perform the integration operation, $RR^{*}$ content of each integral in Eq. \eqref{m23} is written by using the series expansion of each special function \cite{book2},

\begin{equation}
\begin{aligned}
I_{0}(z) &=\sum_{k=0}^{\infty} \frac{\left(\frac{z}{2} \right)^{2k}}{(k!)^{2}}, \\
K_{0}(z) &=-ln\frac{z}{2} I_{0}(z)+  \sum_{k=0}^{\infty} \frac{\left(z \right)^{2k}}{2^{2k}(k!)^{2}} \psi(k+1),   \\
{}_{0}F_{1}(;b;z) &= \sum_{k=0}^{\infty} \frac{\left(z \right)^{k}}{(b)_{k}k!}.
\end{aligned}
\end{equation}

Integrations are performed by first changing the variable of integration according to $u=r-r_{\Delta}$. Since the integrals are to be evaluated near the singularity $(r \rightarrow r_{\Delta})$, the new variable $u$ is very small. This leads us to take $K_{0}(z) \sim -ln\frac{z}{2} I_{0}(z)$ as the dominant term. In addition, the integrand involves the multiplication of power series. The analytic calculation of these integrals becomes possible if we use Cauchy product law of power series as explained below.

\textbf{Definition 1} \textit{(The Cauchy Product of Power Series):} Consider the power series $\sum_{n=0}^{\infty}a_{n}z^{n} $, and  $\sum_{n=0}^{\infty}b_{n}z^{n} $. Consequently, the Cauchy Product of these series can be defined as \cite{book3}

\begin{equation}
\left(\sum_{n=0}^{\infty}a_{n}z^{n} \right)\left(\sum_{n=0}^{\infty}b_{n}z^{n} \right)=\sum_{n=0}^{\infty}\left(\left(\sum_{j=0}^{n}\left(a_{j}b_{n-j} \right)\right)z^{n} \right)=\sum_{n=0}^{\infty}c_{n}z^{n}
\end{equation}
where $c_{n}=\sum_{j=0}^{n}a_{j}b_{n-j} $.

For the sake of simplicity, we take the integration constants $a_{1}$ to $a_{6}$ to be equal to $1$. With the help of the Cauchy product law of power series, integrals can be calculated analytically. \\

First we consider the integral in which the deformation parameter is $0<\gamma<1/2 $. The corresponding square norm becomes

\begin{equation}
\begin{aligned}
\Vert R\Vert ^{2} \sim &\alpha_{1}\sum_{k=0}^{\infty}c_{k}\int_{const}^{0 }\left(ln^{2}\frac{\eta_{1}u^{\gamma^{2}/2}}{2} \right)\left(\frac{\eta_{1}u^{\gamma^{2}/2}}{2} \right)^{k}\frac{du}{u^{2\gamma-\gamma^{2}}}\\
&+ \alpha_{1}\sum_{l=0}^{\infty}c_{l}\int_{const}^{0 }\left(\eta_{2}u^{\gamma^{2}} \right)^{l}\frac{du}{u^{2\gamma-\gamma^{2}}}\\
&-2\alpha_{1}\sum_{n=0}^{\infty}c_{n}\int_{const}^{0 }\left(ln\frac{\eta_{1}u^{\gamma^{2}/2}}{2} \right)\left(u^{\gamma^{2}} \right)^{n}\frac{du}{u^{2\gamma-\gamma^{2}}}
\end{aligned}
\end{equation}
in which $\alpha_{1}=\left(\frac{2p}{1+p} \right)^{4\gamma}\frac{p^{2-2\gamma^{2}}(1+p)^{2\gamma^{2}-1}}{(2mp)^{2\gamma-\gamma^{2}}r_{+}^{2\gamma^{2}-4\gamma-2}}$, $c_{k}=\sum_{l=0}^{k}\frac{1}{\left(l!(k-l)! \right)^{2}}$, $c_{l}=\sum_{t=0}^{l}\frac{1}{(1)_{t}t!(1)_{l-t}(l-t)!}$ and $c_{n}=\sum_{j=0}^{n}\left\{\left(\frac{\eta_{1}}{2}\right)^{j}\frac{1}{(j!)^{2}} \right\}\left\{ \frac{(\eta_{2})^{n-j}}{(1)_{n-j}(n-j)!}\right\}$. \\
The first and third integrals are converging, since $\int_{c}^{0 }x^{k}ln(bx)dx=\frac{c^{1+k} (1-(1+k) \text{ln}[b c])}{(1+k)^2}$  and
$\int_{c}^{0 }x^{k}ln^{2}(bx)dx=-\frac{c^{1+k} (2+(1+k) \text{ln}[b c] (-2+(1+k) \text{ln}[b c]))}{(1+k)^3}$ are finite. The convergence of the second integral can be seen with the help of the comparison test. Since $u$ is very small and positive, we can define the following inequality,

\begin{equation}
u^{\gamma^{2}l-(2\gamma-\gamma^{2})}\leq u^{-(2\gamma-\gamma^{2})}.
\end{equation}
Furthermore, as $\int_{const}^{0 }u^{-(2\gamma-\gamma^{2})}du=\frac{u^{(\gamma-1)^{2}}}{(\gamma-1)^{2}}\left\vert _{const}^{0 }\right < \infty $,  the integral $\int_{const}^{0 }u^{\gamma^{2}l-(2\gamma-\gamma^{2})}du$ is also converging. As a result, when $0<\gamma<1/2 $ the solution is square integrable and the spacetime singularity remains quantum singular. \\

Next, we consider the case when $\gamma>1/2$ but not equal to $1$. The square norm for this particular case becomes

\begin{equation}
\begin{aligned}
\Vert R\Vert ^{2} \sim &\alpha_{2}\sum_{k=0}^{\infty}c_{k}\int_{const}^{0 }\left(ln^{2}\frac{\eta_{3}u^{\gamma^{2}-2\gamma+1}}{2} \right)\left(\frac{\eta_{3}u^{\gamma^{2}-2\gamma+1}}{2} \right)^{k}u^{\gamma^{2}-2\gamma}du\\
&+ \alpha_{2}\sum_{l=0}^{\infty}c_{l}\int_{const}^{0 }\left(\eta_{4}u^{\gamma^{2}-2\gamma+1} \right)^{l}u^{\gamma^{2}-2\gamma}du\\
&-2\alpha_{2}\sum_{n=0}^{\infty}c_{n}\int_{const}^{0 }\left(ln\frac{\eta_{3}u^{\gamma^{2}-2\gamma+1}}{2} \right)\left(u^{\gamma^{2}-2\gamma+1} \right)^{n}u^{\gamma^{2}-2\gamma}du
\end{aligned}
\end{equation}
where $\alpha_{2}=\left(\frac{2p}{1+p} \right)^{4\gamma}\frac{p^{2-2\gamma^{2}}(1+p)^{2\gamma^{2}-1}}{(2mp)^{-2\gamma+\gamma^{2}}r_{+}^{2\gamma^{2}-4\gamma-2}}$. The first and third integrals are very similar to the previous case, and therefore, converging.  The second integral can be calculated via  $\int_{const}^{0 }u^{l(\gamma-1)^{2}+(\gamma^{2}-2\gamma)}du=\frac{u^{(l+1)(\gamma-1)^{2}}}{(l+1)(\gamma-1)^{2}}\vert_{const}^{0}$ and are found as square integrable, which implies that the spacetime singularity remains quantum singular.

Finally, we consider the case when $\gamma=1/2$. In this case, the square norm equation becomes

\begin{equation}
\begin{aligned}
\Vert R\Vert ^{2} \sim &\alpha_{3}\sum_{k=0}^{\infty}c_{k}\int_{const}^{0 }\left(ln^{2}\frac{\eta_{5}u^{1/8}}{2} \right)\left(\frac{\eta_{5}u^{1/8}}{2} \right)^{k}\frac{du}{u^{3/4}}\\
&+ \alpha_{3}\sum_{l=0}^{\infty}c_{l}\int_{const}^{0 }\left(\eta_{6}u^{1/4} \right)^{l}\frac{du}{u^{3/4}}\\
&-2\alpha_{3}\sum_{n=0}^{\infty}c_{n}\int_{const}^{0 }\left(ln\frac{\eta_{5}u^{1/8}}{2} \right)\left(u^{1/4} \right)^{n}\frac{du}{u^{3/4}}
\end{aligned}
\end{equation}
where $\alpha_{3}=\left(\frac{2p}{1+p} \right)^{2}\frac{p^{3/2}(1+p)^{-1/2}r_{+}^{7/2}}{(2mp)^{3/4}}$. As in the previous cases, the first and the third integrals are square integrable and the second integral can be calculated with the help of the comparison test. In doing so, we define the following inequality,

\begin{equation}
 u^{l/4-3/4}  \leq u^{-3/4}
\end{equation}
and since $\int_{const}^{0 }u^{-3/4}du=4u^{1/4}\left\vert _{const}^{0 }\right < \infty$, square integrability is implied and therefore, the spacetime singularity remains quantum singular.

\section*{Data Availability Statement}
No new data were created or analysed in this study.
\section*{Declarations}
\subsection*{Ethical Approval}
Not applicable
\subsection*{Competing Interests}
Not applicable
\subsection*{Author's Contribution}
O.G. and M.H. wrote the article and M.M. did the calculations.
\subsection*{Funding}
Not applicable

\end{document}